\documentclass[epj]{svjour}
\usepackage{amssymb}
\usepackage{graphicx}
\begin{document}
\title{Spurious finite-size instabilities with Gogny-type interactions}
\subtitle{}
\author{M. Martini\inst{1}%
\thanks{\emph{Present address:} martini.marco@gmail.com}
 \and A. De Pace\inst{2}
\thanks{\emph{Present address:} depace@to.infn.it}%
 \and K. Bennaceur\inst{3}
\thanks{\emph{Present address:} bennaceur@ipnl.in2p3.fr}%
}                     
%
%
\institute{DRII-IPSA, 63 boulevard de Brandebourg, 94200 Ivry-sur-Seine, France
\and
Istituto Nazionale di Fisica Nucleare, Sezione di Torino, 
Via P. Giuria 1, I-10125 Torino, Italy
 \and Univ Lyon, Universit\'e Claude Bernard Lyon 1, CNRS, IPNL,\\
UMR 5822, 4 rue E. Fermi, F-69622 Villeurbanne Cedex, France}
\date{Received: date / Revised version: date}
%
\abstract{
Recently, a new parameterization of the
Gogny interaction suitable for astrophysical applications,
named D1M$^{*}$, has been presented. We investigate the possible existence of
spurious finite-size instabilities of this new Gogny force by repeating a study
that we have already performed for the most commonly used parameterizations
(D1, D1S, D1N, D1M) of the Gogny force. This study is based on a
fully-anti\-sym\-me\-trized random phase approximation (RPA) calculation of the
nuclear matter response functions employing the continued fraction technique. 
It turns out that this new Gogny interaction is affected by spurious finite-size
instabilities in the scalar isovector channel; hence, unphysical results are
expected in the calculation of properties of nuclei, like neutron and proton
densities, if this D1M$^{*}$ force is used. The conclusions from this study
are then, for the first time, tested against mean-field calculations in a
coordinate representation for several nuclei.
Unphysical results for several nuclei are also obtained with the D1N
parameterization of the Gogny force.
These observations strongly advocate for the use of the linear response
formalism to detect and avoid finite-size instabilities during
the fit of the parameters of Gogny interactions as it is already
done for some Skyrme forces.
} 
\maketitle

Recently, a new parameterization of the
Gogny interaction suitable for astrophysical applications
has been presented~\cite{Gonzalez-Boquera:2017rzy} since it has been
found~\cite{Sellahewa:2014nia,Gonzalez-Boquera:2017uep} that the most
successful parameterizations of this
force for describing finite nuclei, namely D1S~\cite{Berger:1991zza},
D1N~\cite{Chappert:2008zz} and D1M~\cite{Goriely:2009zz}, commonly suffers from
a rather soft neutron matter equation of state and, as a consequence, are
unable to reach neutron star masses of about $2M_\odot$, as required by recent
astrophysical observations~\cite{Demorest2010,Antoniadis1233232}.

The authors of Ref.~\cite{Gonzalez-Boquera:2017rzy} proposed a
reparameterization scheme that preserves the main properties of the Gogny
force but allows for tuning the density dependence  of the symmetry energy
which, in turn, modifies the predictions for the maximum stellar mass. 
This scheme works well with D1M as a starting point, and led to a new
parameter set, dubbed D1M$^{*}$. The parameters of this new interaction are constrained
by requiring the same saturation density, energy per particle, compressibility,
effective mass  in symmetric nuclear matter, symmetry energy at density
0.1~fm$^{-3}$ as in the original D1M force. The change in the slope is from
$L=24.83$~MeV, the value of D1M,  to $L=43.18$~MeV.
This latter value makes D1M* the first Gogny-type interaction
able to predict neutron star masses in agreement with the already mentioned
recent observations.

In order to preserve the good performance of D1M in describing nuclear structure
features of  finite nuclei the authors of Ref.~\cite{Gonzalez-Boquera:2017rzy}
checked that the basic bulk properties of D1M$^*$, such as binding energies
and charge radii of even-even nuclei, remain globally unaltered as compared
to D1M. For this purpose, they carried out Hartree-Fock-Bogolyubov (HFB)
calculations for 620 even-even nuclei of the 2012 AME~\cite{Audi12} database
using the~\mbox{HFBaxial} code~\cite{HFBaxial} which solves the HFB equations
in a harmonic oscillator basis.

Beyond these evaluations performed by the authors, it is interesting to check
the behavior of this newly adjusted interaction with respect to the development
of finite-size instabilities, since it was pointed out that they can be hidden
by the use of a representation of the quasiparticle wave functions on a
limited number of harmonic oscillator shells~\cite{Hellemans:2013bza}
whereas these instabilities can develop when the calculation is performed
on a mesh.

To study the possible onset of finite-size instabilities,
we start by repeating here the same study as the one published
in Ref.~\cite{DePace:2016lwp} where we developed a fully-anti\-sym\-me\-trized
random phase approximation (RPA) calculation of the nuclear matter response
functions based on the continued fraction (CF) technique to investigate the
possible existence of spurious finite-size instabilities of the Gogny forces.
In Ref.~\cite{DePace:2016lwp} we considered the most commonly used
parameterizations of the Gogny force (D1~\cite{Gogny:1973}, D1S, D1N, D1M),
as well as recent generalizations that include tensor terms. Here we complete
the study by considering D1M$^{*}$. Since finite-size instabilities can easily
be revealed with mean-field calculations on a mesh~\cite{hfbrad,ev8,finres},
a series of such HFB calculations is then performed to possibly confirm their
appearance for some of the considered interactions.

\begin{figure*}
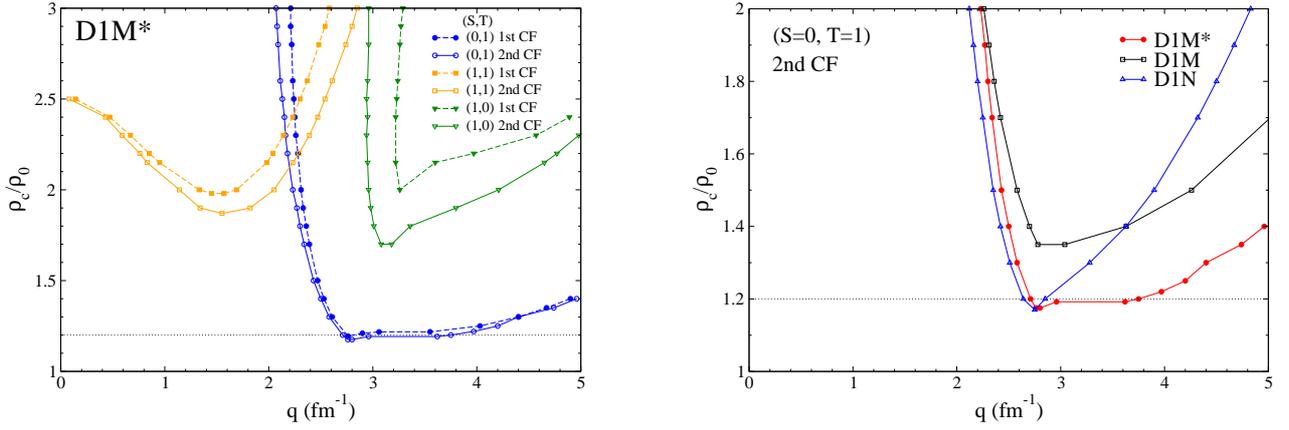

\begin{center}
\begin{tabular}{cc}
\includegraphics[clip,width=0.425\textwidth]{fig_rhoc_vsq_d1mstar.eps}
~~~~&~~~~
\includegraphics[clip,width=0.425\textwidth]{fig_rhoc_vsq_01_2cf_d1nmmstar.eps}
\end{tabular}
\end{center}
\caption{Left panel: Critical densities $\rho_c$ divided by the constant
  value corresponding to the saturation density $\rho_0=0.16$ fm$^{-3}$ as a
  function of the
  transferred momentum $q$ (in fm$^{-1}$) for the D1M$^{*}$ 
  parameterization of the Gogny force. The calculations of
  $R_{(S,T)}(q,\omega=0)$, through which the critical densities are deduced,
  are performed at first and second order in the continued fraction
  expansion. Right panel: Critical densities as a function of the transferred
  momentum in the scalar-isovector channel for the D1M$^{*}$, D1M and D1N
  parameterizations of the Gogny force. The calculations are performed at
  second order in the continued fraction expansion.
  To guide the eye, the value $\rho_c/\rho_0=1.2$ is also plotted.
}
\label{fig_rhoc}
\end{figure*}

The key quantities to investigate unphysical finite-size instabilities are the
critical densities $\rho_c$, \textit{i.e.} the lowest densities at which the
nuclear response calculated at zero transferred energy, $R_{(S,T)}(q,\omega=0)$,
exhibits a pole. On the left panel of Fig.~\ref{fig_rhoc} we show the critical
densities in the three
spin-isospin $(S,T)$ channels $(0,1)$, $(1,1)$ and $(1,0)$ as a function of the
transferred momentum $q$ for the D1M$^{*}$ force. We disregard the $(S,T)=(0,0)$
channel and its corresponding physical  spinodal instability. All the details
of the calculations are here omitted since they can be found in
Ref.~\cite{DePace:2016lwp}. The only difference with respect to this
article being the parameters of the considered Gogny forces.

The result that we obtain in the scalar-isovector channel $(S,T)=(0,1)$ is
particularly important: the critical density rapidly decreases with $q$
reaching values
around $\rho_c\simeq1.2 \rho_0$, where $\rho_0=0.16$ fm$^{-3}$ is the empirical
saturation density. Furthermore the values of $\rho_c$ only slowly increase
with $q$ after the minimum.
In Ref.~\cite{Hellemans:2013bza}, a systematic quantitative analysis of the
connection between finite nuclei and nuclear matter instabilities in the
$(S,T)=(0,1)$ channel was performed finding that a functional is stable if
the lowest critical density at which a pole occurs in nuclear matter
calculations is larger than the typical central density often obtained
for $^{40}$Ca which is, in practice, around 1.2 times the saturation density.
In addition, one also has to verify that this pole represents a distinct
global minimum in the $(\rho_c,q)$ plane. It therefore appears that neither of
the two stability criteria to avoid spurious finite-size instabilities
established in Ref.~\cite{Hellemans:2013bza} is satisfied for the
new D1M$^{*}$ force.  The behavior of the critical density in the $(S,T)=(0,1)$
channel for D1M$^{*}$ is worse with respect to the corresponding behavior for
D1M and D1N, as one can observe on the right panel of Fig.~\ref{fig_rhoc}.

One of the conclusions of the analysis of Ref.~\cite{DePace:2016lwp} was that
the D1N parameterization of the Gogny force should be treated with some caution
since the stability criteria are not strictly respected by this force.

\begin{figure}
\begin{center}
\includegraphics[clip,height=0.4\textwidth,angle=270]{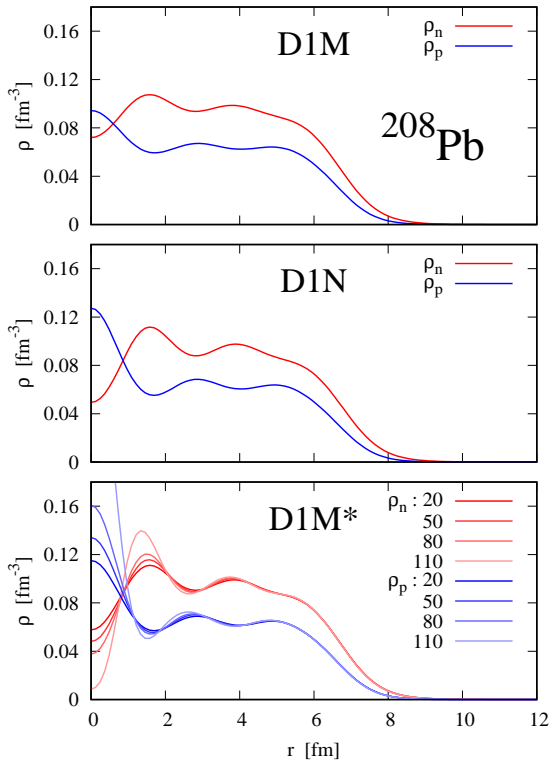}
\caption{Neutron (in red) and proton (in blue) densities
  obtained from Hartree-Fock calculations for $^{208}$Pb with the interactions
  D1M (top panel), D1N (central panel) and D1M* (bottom panel). Since the
  interaction D1M$^*$ does not lead to a self-consistent convergent solution,
  different levels of red and blue are used to plot the densities after
  different numbers of iterations as indicated on the figure.
  \label{fig_208Pb}}
\end{center}
\end{figure}

\begin{figure}
\begin{center}
\includegraphics[clip,height=0.4\textwidth,angle=270]{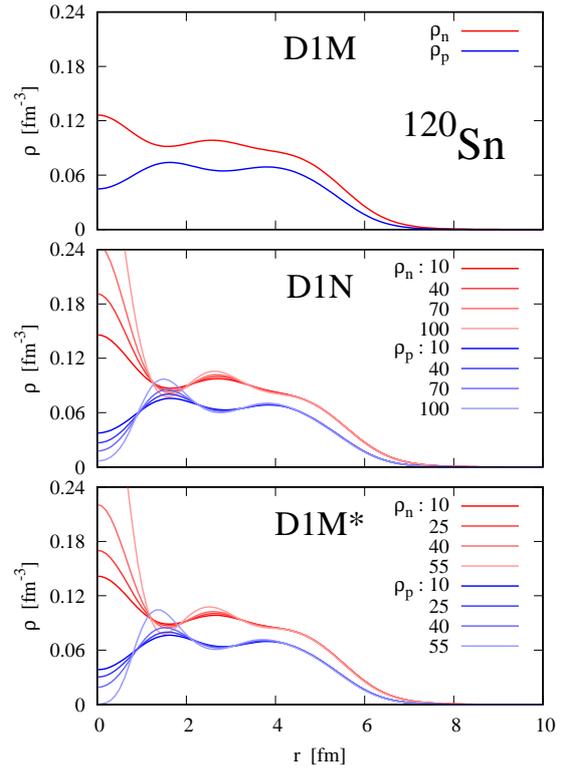}
\caption{Same as figure~\ref{fig_208Pb} for $^{120}$Sn.
  In this case, the calculations did not converge using D1N or D1M$^*$
  for which different levels of red and blue are used to plot the
  densities after different numbers of iterations as indicated on the figure.
  \label{fig_120Sn}}
\end{center}
\end{figure}

To possibly confirm the appearance of finite-size instabilities in nuclei
calculated with the D1M* force and to remove the ambiguity concerning D1N
we have performed a series of mean-field calculations using the recently
developed code {\tt FINRES}$_4$~\cite{finres} which solves the
self-consistent Hartree-Fock-Bogolyubov equations on a mesh for finite-range
interactions using the method described in~\cite{Hooverman:1972}.
This series of calculations is done for three very different
nuclei: the doubly-magic stable nucleus $^{208}$Pb, the semi-magic nucleus
$^{120}$Sn for which pairing is active,
and the very light symmetric nucleus $^{4}$He. This latter is not
necessarily supposed to be correctly described by a mean-field calculation,
but it represents an interesting test case for the study of
instabilities in the $(S,T)=(0,1)$ channel because of its small number
of constituents and its approximate proton-neutron symmetry only weakly
broken by the small Coulomb field.

The calculations were done in a spherical box with radius $R=20$~fm on a mesh
with a spacing $\delta r=0.1$~fm and with an expansion up to
$\ell_\mathrm{max}=19$
for the densities. The Coulomb exchange contribution and the center-of-mass
correction were treated exactly. With the D1M interaction, the
self-consistent calculations were initialized from a schematic Woods-Saxon
potential. The final converged densities obtained with D1M were then
used to initialize the calculations for D1N and D1M*.

On Figure~\ref{fig_208Pb} are represented the proton and neutron densities
from an Hartree-Fock calculation with the D1M, D1N and D1M* Gogny interactions
for $^{208}$Pb. With the D1M* interaction, the calculation did not converge and
led to oscillations of the isovector density with very large amplitude. The
proton and neutron densities for different numbers of HFB
iterations are represented on the figure. The calculations converged with D1M
and D1N, for which only the final densities are represented.
We can nonetheless observe that the difference between the proton and neutron
densities at $r=0$ is much larger with D1N than with D1M. This large
difference is due to significant oscillations of the densities in the bulk
of the nucleus and can be interpreted as a warning signal for a finite-size
instability close to show up.

A series of HFB calculations  were also done for $^{120}$Sn with the same
interactions. The corresponding densities are represented on
Fig.~\ref{fig_120Sn}. In this case, we observe that the calculation
converges with D1M, but neither with D1M* nor D1N. For the latter two
interactions, in the same manner as on Fig.~\ref{fig_208Pb}, we
have plotted the proton and neutron densities for different number
of iterations during the calculations.

From these calculations we can conclude that D1N does not always
lead to convergent results when it is used to calculate finite nuclei
on a mesh. The non-convergence observed here with $^{120}$Sn is not the
only one we encountered. Similar instabilities were observed for
nuclei with moderate or large neutron number and for calculations
done with the HF or HFB approximations
(for example $^{60}$Ca calculated with the HF approximation).

On the right panel of Fig.~\ref{fig_rhoc}, we observe that for D1N,
the critical densities
in the $(S,T)=(0,1)$ channel approach the saturation density for a relatively
narrow interval of transferred momenta $q$. Since the typical wave-length
of the oscillations leading to an instability is $\lambda\sim 2\pi/q$,
we can expect that D1N cannot lead to the appearance of instabilities
in very small nuclei. The situation is rather different with D1M*
which shows a plateau of critical densities at approximately
$1.2\times\rho_\mathrm{sat}$ for $2.8~\mathrm{fm}^{-1}\lesssim q\lesssim
4~\mathrm{fm}^{-1}$. This means that with D1M*, finite-size instabilities
may well appear in relatively small nuclei.

\begin{figure}
\begin{center}
\includegraphics[clip,height=0.4\textwidth,angle=270]{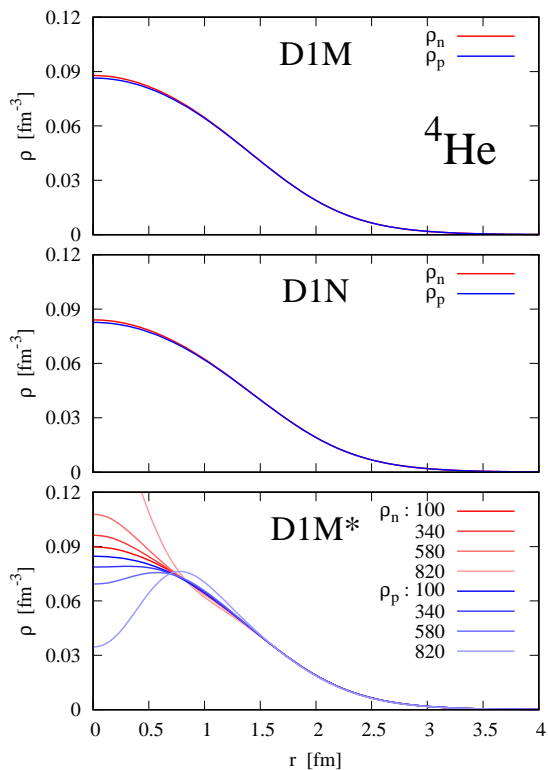}
\caption{Same as figure~\ref{fig_208Pb} for $^{4}$He.
  For D1M$^*$, different levels of red and blue are used to plot the
  densities after different numbers of iterations as indicated on the figure.
  \label{fig_4He}}
\end{center}
\end{figure}

To test this conjecture, we have performed a series of calculations for
$^{4}$He. The results are presented on Fig.~\ref{fig_4He}.
As for $^{208}$Pb and $^{120}$Sn, the interaction D1M leads to a convergent
result. As expected from the aforementioned argument, the calculation
did converge with D1N but not with D1M* for which, despite the very
limited number of nucleons and the tiny asymmetry of the system,
an instability develops.

Let us mention that not all nuclei lead to finite-size instabilities
with D1M*. For example, we obtained convergent results with D1M* for
$^{16}$O, $^{100}$Sn and $^{176}$Sn (and with all $N = Z$ nuclei calculated
without Coulomb interaction).
Let us also mention that random tests with
D1S for nuclei with mass number above 40 did not show any sign of
instabilities.

Before we conclude, we can comment on the results obtained in
Ref.~\cite{Gonzalez-Boquera:2017rzy} using the \mbox{HFBaxial} code.
The convergence of the calculations, with results in good agreement with data,
obtained by the authors for finite nuclei are owed to the use of a basis
with a given number of oscillator shells. This representation strongly
renormalizes the interaction and inhibits the development of instabilities
and therefore makes the obtained results sound. But this renormalization
has several important drawbacks.

\begin{figure}
\begin{center}
\includegraphics[clip,height=0.45\textwidth,angle=270,viewport=40 80 530 730]{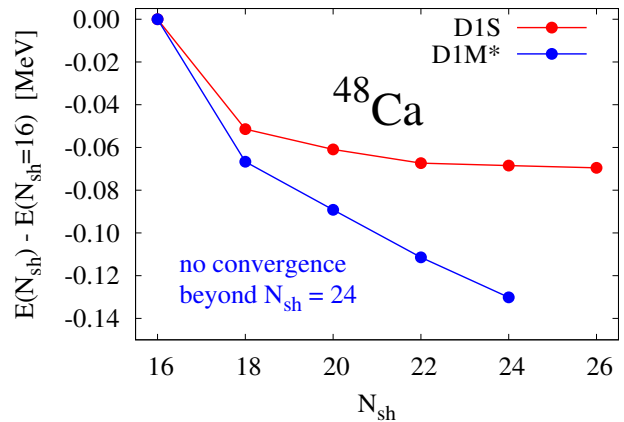}
\caption{Binding energy of $^{48}$Ca as function of the number of shells
         $N_\mathrm{sh}$ calculated with the interactions D1S and D1M*
         using the code HFBTHO. The result $E(N_\mathrm{sh}=16)$ is taken as
         a reference in both cases. We were not able to obtain a converged
         result beyond $N_\mathrm{sh}=24$ with D1M*.\label{cv48ca}}
\end{center}
\end{figure}

First, the D1M$^*$ (or D1N) interaction
should only be used with the basis employed to fit its parameters.
This, of course, prevents its use for calculations on a mesh, but also
leads to results that depend on the size of the basis when the
equations are solved by expanding the solutions on harmonic oscillator
wave functions. This is exemplified on Figure~\ref{cv48ca} where
the binding energy $E(N_\mathrm{sh})$ for $^{48}$Ca is plotted as a function
of the number of shells $N_\mathrm{sh}$ for the interactions D1S and D1M* using
the code HFBTHO~\cite{PEREZ2017363}. On can clearly see that $E(N_\mathrm{sh})$
rapidly reaches a plateau with D1S while, with D1M*, its change is
significant for each increase of $N_\mathrm{sh}$ up to $N_\mathrm{sh}=24$,
the last value with which we were able to obtain a converged results
using HFBTHO.

Second, the strong renormalization on the harmonic oscillator basis
breaks the link with the properties of the interaction in infinite
nuclear matter where the properties of the saturation point and
the various equations of state are calculated without accounting for
this renormalization but with the assumption that local densities
are constant in space.

In this article, we have shown that it exists a robust correspondence between
the finite-size instabilities which can be inferred from RPA calculations in
infinite nuclear matter and the one observed in nuclei for Gogny forces as
it is the case for Skyrme interactions~\cite{Hellemans:2013bza}.
Therefore we encourage to use infinite nuclear matter RPA calculations
as a tool to avoid these instabilities. This would allow to use
the obtained parameters sets on any basis and would maintain the link
between infinite nuclear matter and finite nuclei. Such a tool already
exists~\cite{DePace:2016lwp} and could be used in the fit protocol for
the construction of new Gogny forces, as already done for the Skyrme
functionals \cite{Kortelainen:2013faa,pastore2013,jodon2016}. 

\section*{Acknowledgments}

The authors are grateful to D. Davesne and J. Meyer
for valuable discussions during the development of this study.

\bibliographystyle{epj}

\bibliography{paper_instab}

\end{document}